# A 3-step Low-latency Low-Power Multichannel Time-to-Digital Converter based on Time Residual Amplifier

Florent Bouyjou[1], Eric Delagnes[1], Fabrice Couderc[1], Damien Thienpont[2], José David Gonzalez-Martinez[2], Frederic Dulucq[2], Fabrice Guilloux[1] and Irakli Mandjavidze[1]

*Abstract*— This paper proposes and evaluates a novel architecture for a low-power Time-to-Digital Converter with high resolution, optimized for both integration in multichannel chips and high rate operation (40 Mconversion/s/channel). This converter is based on a three-step architecture. The first step uses a counter whereas the following ones are based on two kinds of Delay Line structures. A programmable time amplifier is used between the second and third steps to reach the final resolution of 24.4 ps in the standard mode of operation. The system makes use of common continuously stabilized master blocks that control trimmable slave blocks, in each channel, against the effects of global PVT variations. Thanks to this structure, the power consumption of a channel is considerably reduced when it does not process a hit, and limited to 2.2 mW when it processes a hit. In the 130 nm CMOS technology used for the prototype, the area of a TDC channel is only 0.051 mm². This compactness combined with low power consumption is a key advantage for integration in multi-channel front-end chips. The performance of this new structure has been evaluated on prototype chips. Measurements show excellent timing performance over a wide range of operating temperatures (-40°C to 60°C) in agreement with our expectations. For example, the measured timing integral nonlinearity is better than ±1 LSB (±25 ps) and the overall timing precision is better than 21 ps RMS.

## I. INTRODUCTION

Time-to-digital converters (TDCs) are widely used in High-Energy Physics and in many other applications that require highly precise time measurements. During the last few years, fast timing detectors have become more and more important as the extra information they provide open new possibilities for data filtering or particle identification. Today, one of the main challenges of the CERN collaborations for the High Luminosity LHC (HL-LHC) era [1, 2] will be the large pileup with up to 200 simultaneous proton collisions per bunch crossing to record in a detector. The most efficient way to mitigate the pileup in the most occupied regions of detectors will be to use the timing information, opening the way to 4D-tracking (spatial and temporal). For this purpose, a timing precision in the range of 30 ps RMS is required. Outside the field of HEP, these new high-precision timing detectors also pave the way for improvements or new concepts in medical applications (TOF-PET…) [3] or instrumentation (LIDAR) [4].

Today, a large spectrum of TDC architectures is available for integration in ASICs and Field-Programmable Gate Arrays (FPGAs) [5]. Most of them are based on digital delay line (DL). The most advanced DL-based TDC ASICs exhibit timing resolution in the 10 ps range and the best ones reach a few ps [6, 7], using for example the resistive interpolation technique. But, this is achieved only after complicated calibration procedures, with the penalties of high power consumption, low rate operation, or large area, incompatible with integration in multichannel high-speed front-end chips.

For our target requirements (few 10 ps precision range and 40 Mconversion/s) more moderate power consumptions have been recently achieved using multistep structures based on Vernier [8] or Time Amplifier (TA) methods for the last stage. Modern TA structures are able to amplify time intervals as small as few tens of ps by an order of magnitude so that they become measurable by standard methods. This makes it a very attractive technique for fast, high-resolution TDCs [7]. However, TAs are generally very sensitive to PVT (Process, Voltage, and Temperature parameters) variations. Indeed, the gain of TAs is usually set by open-loop high-speed elements, depending highly on PVT parameters.

This paper presents the design of a 3-step low-power multichannel TDC, inspired by [9], based on time residue amplification but with extended dynamic range and improved robustness against PVT variations. The first step catches the output of a counter operating at 160 MHz. The second step, using a DL, refines the time-stamping within the 160 MHz period, with a 200 ps resolution. The final resolution of few tens of picoseconds is obtained by the third step that uses a DL fed by a programmable Pulse-Train Time Amplifier optimized for a large dynamic range and an enhanced stability.

---

[1] The authors are with IRFU, CEA, Université Paris-Saclay, F-91191 Gif-surYvette, France. [2] The authors are with Ecole Polytechnique, CNRS/IN2P3, Omega, Palaiseau, France.
E-mail: florent.bouyjou@cea.fr, eric.delagnes@cea.fr, fabrice.couderc@cea.fr

This work was supported in part by the P2IO Labex P2IO LabEx (ANR-10 LABX-0038) in the framework « Investissements d'Avenir » (ANR-11 IDEX-0003-01) managed by the French National Research Agency (ANR).

The paper is organized as follows. Section II describes the overall architecture of the 3-step multichannel TDC. Section III presents in details the main building blocks used in the TDC. The implementation and the experimental results are released in Section IV. Finally, Section V draws the conclusions and outlines the future work.

## II. DESCRIPTION OF THE THREE-STEP TDC

### A. Requirements

Our goal was to design a multichannel TDC IP block, fulfilling the CMS HGCAL [10] requirements, but also easily reusable for other applications, implemented in the 130 nm CMOS technology from TSMC widely used by the HEP community. Its resolution must be better than 30 ps over a range of at least 25 ns (>10-bit dynamic range), corresponding to the LHC bunch crossing period, and its target timing precision should be of 20 ps RMS (after calibration). Optional features of the design allow to extend its dynamic range, but the paper is focused on the nominal mode of operation (10-bit range with 25 ps LSB). The total conversion time must be less than 25 ns leading to a conversion rate of up to 40 Mconversion/s required for the LHC operation. Because this device will be used to read cooled silicon detectors, it is designed to operate at temperatures as low as −30°C. But because it will be tested at room temperature and intended for a wide range of applications, its operating range extends to 65°C. Finally, as this TDC will be used in a detector with several tens of million channels, we have chosen to minimize the required calibration procedures as well as the number of calibration parameters to be determined, stored and used to reach its operating performance.

### B. Global architecture

Most modern high resolution fast TDCs are based on DL made of cascade of elementary digital delays (DE). In an analogous way to the structure of flash ADCs, one could imagine using a $2^N$ DE delay chain to encode the full range of an N-bit TDC. Of course, this would be at the cost of a large silicon area and for large dynamic range, the accumulated jitter in the delay line would degrade the precision. To avoid this, the proposed TDC is based on the 3-step structure shown on Fig. 1.

The TDC is driven by a 160 MHz (CLK) (6.25 ns period $T_{ck}$) generated by an on-chip 1.28 GHz PLL block, locked on an external 40 MHz reference clock. This clock sequences an unique 8-bit Gray counter which outputs are broadcasted to all the TDC channels integrated within the chip. When an event occurs on a channel, the counter output is captured in a local 8-bit register. Only the last 2 bits are used for the nominal use case (25ns coding range), the 6 most significant bits can be used to extend the range up to 1.6 μs or to provide the redundancy required to merge the data from the TDC with a timestamping performed elsewhere at lower frequency. To refine the measurement, a Coarse TDC (CTDC) based on a 32-step DL (CDL) provides 5 more bits. The small length of the CDL limits the jitter accumulation along the delay chain and its 195 ps step is easily achieved in the technology node we use. To extract the less significant bits, the time residue between the hit occurrence and the next step of the CDL is multiplied by 8 (optionally by 16) by a time amplifier (TA) before being coded by a Fine TDC (FTDC) over 3 (optionally 4) bits using another DL (FDL). Then, a digital block decodes and combines the data from the Gray counter, the CTDC and the FTDC to form the TDC data coded over up to 17 bits (up to 1.6 μs range with 12.5 ps step). But, for the nominal ToA (Time of Arrival) mode of operation in our primary application in HGCAL, only 10 bits are kept (25ns with 25 ps step).

The last operations starting from the time amplification are totally asynchronous, but the whole TDC block has been designed to guarantee that complete conversion time will not exceed 25 ns.

In practice, a central common block integrates some reference voltage generators, the PLL, the Gray-counter timebase and the master DLLs used to stabilize the in-channel CDLs and FDLs. This block generates and broadcasts signals to "slave blocks" localized in every channel. Each slave block integrates a register to capture the Gray counter state, a CDL, a TA, a FDL and the associated coding logic. An integrated asynchronous clock generator (ACG) allows to optionally trigger channels, selectable by slow control, with a tunable frequency periodic signal asynchronous to the TDC clock. This block will be used for *in-situ* testing, characterization or calibration of the TDCs with statistical methods without the need of extra component.

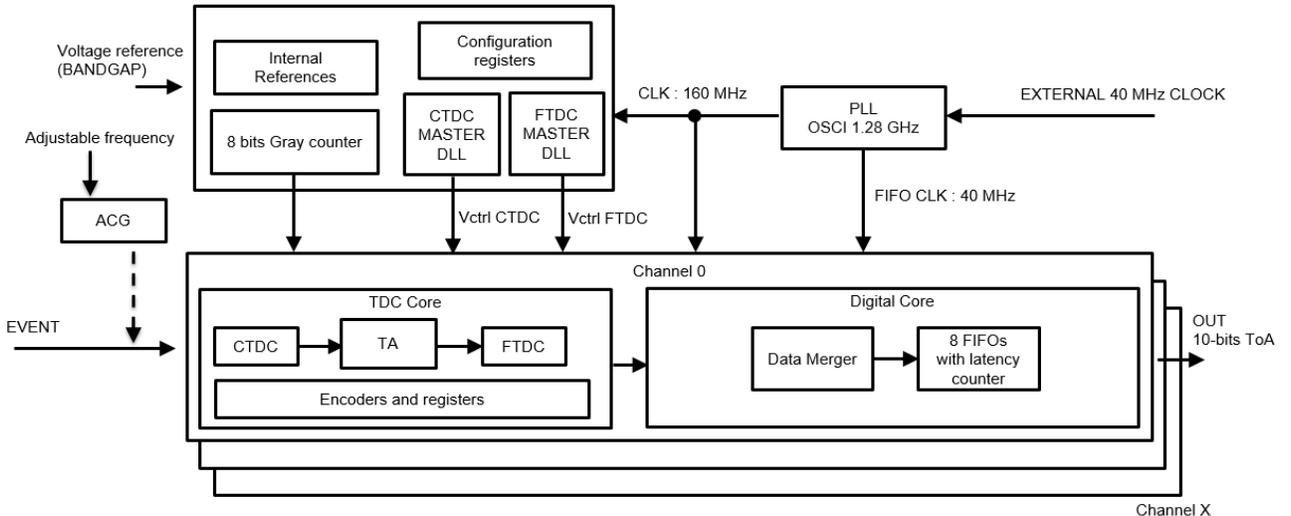

Fig. 1. Block-diagram of the novel 3-steps multi-channel TDC.

To limit the single event effects due to highly ionizing particles, the most critical parts of the design, such as the configuration registers, the Gray counter and the state machines, are triplicated.

### III. DETAILED DESCRIPTION OF THE MAIN BUILDING BLOCKS

#### A. *Counter, event synchronization and associated memorization*

The most significant bits of the TDC are produced by a Gray counter common to all TDC channels running at 160 MHz. Its 8-bit output is broadcasted to the channels. When an event is detected in the front-end part of a channel, the output of its discriminator (*EVENT*) rises to '1'. A latched version of this signal (*HIT*) is provided by a DFF (delay flip-flop). Then, the *HIT* is synchronized by the 160 MHz CLK to produce the *FLASH* signal. To reduce metastability effects, this is achieved by two DFFs put in series, respectively driven by the *CLK* and the *CLK* delayed by about 800 ps by 4 DEs identical to those used in the CDL, each with a $\tau_Q$ delay.

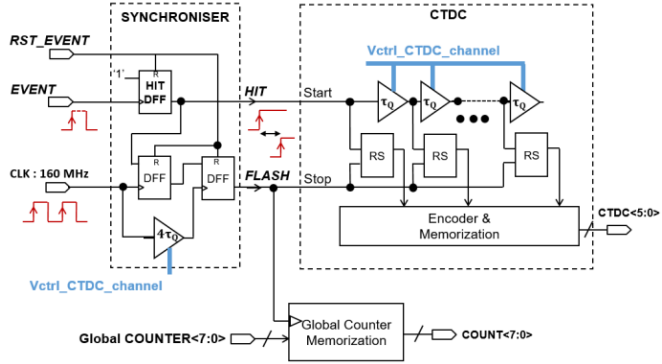

Fig. 2. Block-diagram for the two first stages of the TDC and the event synchronizer.

As shown on Fig. 2, the *FLASH* signal is first used to capture the state of the common counter. Following the Nutt method [11], [12], [13] this *FLASH* signal is also sent as a *Stop* signal to the second stage of the TDC that will measure the time difference between its arrival and the arrival of the asynchronous HIT, used as a *Start*. Thus the time range to code with the second stage of the TDC is comprised between $4\tau_Q$ and $T_{ck} + 4\tau_Q$. To secure the operation the *HIT* DFF is only released at the very end of the conversion by the *RST_EVENT* signal. This implies that only one hit can be processed during a 25 ns period.

#### B. *2nd stage: delay-line coarse TDC and master DLL*

##### 1) *Introduction*

One of the most popular digital methods of Time to Digital conversion for low power multi-channel applications is based on the use of DL. This type of design makes it easy to achieve a time accuracy of a few hundred of picoseconds with a fairly simple and compact structure. In addition, it can be almost completely inactive when not triggered, allowing for very low power operation.

As shown on Fig. 2, our DL-CTDC consists of a chain of N identical DE with a propagation time $\tau_Q$ tunable by a $V_{ctrl}$ voltage. The first delay is fed by a *Start* signal. A memory element, which is a RS latch in our design, is associated to each delay tap to capture the state of the DL when the *Stop* signal arrives. The encoded information from the RS latches is a measure of the delay between the *Start* and the *Stop* signals. The range of the DL is defined as the product of $\tau_Q$ by N, the number of cascaded delay cells, and its resolution is set by the minimum achievable delay and thus depends on the IC technology. Actually, the effective measurement precision is affected by the stability (with temperature, radiation and power supply), the reproducibility (of process parameters) and the uniformity of the delays. Moreover, it can be shown, that each tap of the DL accumulates random jitter proportionally to $\sqrt{N}$ (or N in case of coherent noise source). For this reason, we have chosen to limit the number of taps. To cover the range corresponding to a $T_{ck}$ duration, 32 taps of 195.3 ps are used and 10 extra taps are added to give some range margin allowing to compensate for parasitic delays that will be measured as offsets. This also brings some redundancy allowing to merge the data from the CTDC with those of the counter without any ambiguity. As a consequence the CTDC will produce 6 bits, among which 5 are really significant.

##### 2) *Delay Line Loop and Master DLL*

DE used in the DL are usually sensitive to process parameter variations, temperature, voltage and irradiation. To improve the stability of a DL, a popular solution consists in locking the total delay of a delay chain to a reference time using a delay-locked loop (DLL) structure [14]. For this purpose, as shown on Fig. 3, a reference signal (*refMDLL*) is continuously compared to the output of the DL (*outMDLL*) obtained from its input (*InMDLL*) delayed by a reference value $T_{ref}$. This comparison is achieved by a phase-frequency detector (PFD) controlling a charge pump (QP), the output of which is low-pass filtered to generate the control voltage (*Vctrl*) tuning the duration of the individual delays. In our design, the *InMDLL* signal is provided by a programmable frequency divider sequenced by the 160 MHz CLK. The same signal is delayed by a DFF clocked by the same 160 MHz signal to generate *refMDLL* so that $T_{ref}$ is equal to $T_{ck}$.

As the total delay of the DLL is locked to $T_{ck}$, the average delay of each delay step will be $\tau_Q = T_{ck}/N$.

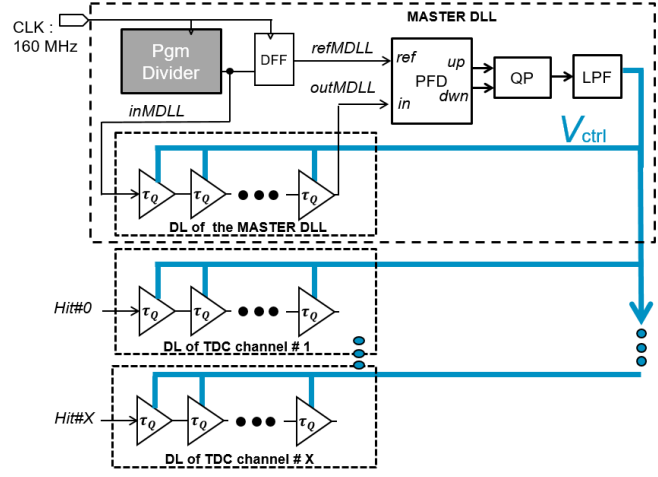

Fig. 3. Block diagram of the MASTER DLL that provides the DL control voltage to the TDC "slave" channels in order to overcome PVT variations.

In practical implementation, it is difficult to use this system for every channel of multichannel TDC without introducing any dead time due to dedicated calibration phases. But, as shown on Fig. 3, such a servo-controlled structure can be used for a master channel only producing the control signal $V_{ctrl}$ for all the "slave" DLs in channels, which are used for measurements. The reproducibility of the measurements then relies on the matching of the slave delay line delays with those of the master DL, so that the two kind of delay lines have to be strictly identical in order to have the same temporal response for PVT variations. Practically, we have chosen to use a master DLL for a group of 36 slave CDLs so that in our prototype 2 master DLLs are controlling 72 channels.

For a DL (master or slave channel), the theoretical total delay is defined as:

$$T_N = N\,\tau_Q = NK_{VDL}V_{ctrl}, \qquad (1)$$

where $K_{VDL}$ is the delay transfer function of a DL element in $s.V^{-1}$.

Usually the voltage-to-delay transfer-function of a DE is not linear. Nevertheless, for simplicity, in the next sections we will consider $K_{VDL}$ to be constant on a reduced range of control voltage.

### 3) Slave DLs and fine delay trimming

Ideally, in our design, $T_N$ should be equal to $T_{ck}$. Practically, the DEs are suffering from mismatches that cause delay spread for a fixed control voltage [14]. For this reason, each step $i$ along the CDL of the channel $c$ exhibits a positive or negative delay error $\varepsilon_{c,i}$ [15-16] and for the DLL the total accumulated delay including errors, $T_{N\varepsilon}$, is given by the following equation:

$$T_{N\varepsilon} = N\tau_Q + \sum_{i=1}^{N} \varepsilon_{c,i}. \qquad (2)$$

For the Master DLL, the closed-loop forces the total DL delay to be locked to $T_N = T_{ck}$ regardless of any PVT variation so that $\sum_{i=1}^{N} \varepsilon_i = 0$. This is not ensured for the slave DLs, which are not servo-controlled, so that $T_{N\varepsilon}$ can be smaller or larger than $T_{ck}$. This can lead to a gap or an overlap problem between the CTDC and the counter ranges, which may drastically affect the overall differential non-linearity (DNL) of the TDC to the point of generating missing codes.

If we consider, that all the TDCs of a multi-channel chip (here 72 channels in two groups of 36 with the same MASTER DLL) should have a DNL better than $\pm 1$ LSB $= \pm 24.4$ ps and if we want to limit the yield rejection of chips to 2% because of this criteria, this implies:

$$\left[\text{erf}\left(\frac{\tau_Q}{\sqrt{2} \cdot \sigma_D}\right)\right]^{72} < 0.98 \text{ requiring } \sigma_D < 6.7 \, ps \, RMS \,. \qquad (3)$$

Where $\sigma_D$ quantifies the spread over the channels of the total duration $T_{N\varepsilon}$ of the DL due to mismatch.

A Monte-Carlo simulation of DL with reasonably sized transistors, compatible with a compact design, leads to estimate $\sigma_D = 28 \, ps \, RMS$. This is 4 times larger than the requirement and even larger than the value of a LSB. A simple design optimization cannot solve this issue, so that we have chosen to adjust the total delay of each channel.
From (1) and (2) we can calculate:

$$T_{N\varepsilon} = N\tau_Q + \sum_{i=1}^{N} \varepsilon_{n,i} = N \, K_{VDLs}(1 + \varepsilon_{KVDL})V_{ctrl} \qquad (4)$$

where $K_{VDLs}$ is the stabilized value of $K_{VDL}$ thanks to the master DLL and $\varepsilon_{KVDL}$ is the mean relative deviation from the foreseen transfer function (or gain) of the DE of the slave DL.

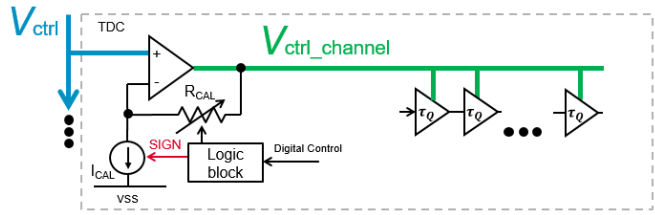

Fig. 4. Block diagram of the per-channel Delay-Line gain correction.

To cancel the error term, for each channel, we adjust slightly the control voltage by adding a small voltage-shift $V_s$,

$$V_s = -\frac{\varepsilon_{KVDL}}{1 + \varepsilon_{KVDL}}V_{ctrl} \,, \qquad (5)$$

to the $V_{ctrl}$ voltage to bring back the total DL delay to its expected value :

$$T_N = N \, K_{VDLs}(1 + \varepsilon_{KVDL})(V_{ctrl} + V_s) \,. \qquad (6)$$

Fig. 4 shows the practical implementation of this correction applied in each channel. It uses an Operational Transconductance Amplifier (OTA) with an adjustable resistor $R_{CAL}$, in the feedback loop through which flows a current $I_{CAL}$. The "Logic block" allows programming the value of $R_{CAL}$ and the direction of the $I_{CAL}$ current. The corrected control voltage is given by:

$$V_{ctrl\_channel} = V_{ctrl} + I_{CAL}R_{CAL} + V_{OFF} \,, \qquad (7)$$

where $V_{OFF}$ is the offset error of the AOP (less than 3 mV RMS). Note that this offset value will also be compensated by the calibration system. Optionally, $R_{CAL}$ and $I_{CAL}$ can be set to 0 so that the AOP only copies $V_{ctrl}$. The design of this block has been optimized to reduce its sensitivity to PVT using references connected to the chip's internal bandgap reference.

### C. 3r$^d$ stage: time amplifier and fine gated delay-line TDC

#### 1) Introduction
This third stage improves the resolution provided by the CDL (195 ps) by a factor of 8 (optionally 16) by measuring finely the timing of the hit signal inside a CDL element.
For this purpose, first a residue generator block provides "residue pulse" with a duration related to the time interval between the *HIT* signal and the next tap of the CDL. Then we amplify the duration of the "residue pulse" using a Time Amplifier (TA). This "amplified delay" is then measured with a Fine Gated Delay Line (FGDL) TDC.

#### 2) Concept and design of the TA
The TA used in the proposed architecture is an improvement of the Pulse-Train amplifier introduced in [9], which generates a train of $A$ copies from an input pulse (*IN_TA*) of $T_R$ duration.
The design, shown on Fig. 5 consists of a ring of M non-inverting delay ($\tau_d$) elements associated with a counter. Between the arrivals of events, the inputs of the delay elements are forced to '0'. When an event occurs, the output of the first delay is set to '1' during the input pulse duration. As long as the *TA_EN* signal is high, the generated pulse is propagated in the ring. As the total propagation of the loop, $T_{loop} = M\tau_d$, is longer than the pulse's duration, the loop generates a train of pulses, similar to the input ones, spaced by $T_{loop}$. This process is stopped when the counter, counting the generated pulses, reaches the programmed value $A$ and inhibit the TA by resetting the *TA_EN* signal. The total duration of the state '1' in the output train of pulses generated at the TA output, named "pulse-train duration" hereafter, corresponds to the duration of the input pulse multiplied by $A$. This design consumes power only during the pulse multiplication process, which is a clear advantage for integration into a low power chip.

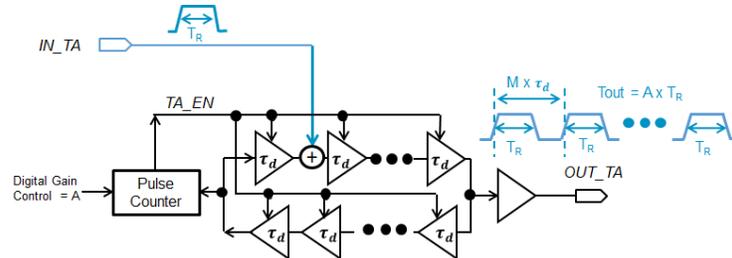

Fig. 5. Block-diagram of the pulse-train time amplifier based on M delay looped cells and a pulse counter

With such an architecture, there is a risk to loose very short residue pulses during the circulation operation. This would clip the transfer function of the TA resulting in large nonlinearities of the TDC response. To avoid this, the input pulse of the TA, whose rising edge is defined by the *HIT* signal, falls back only on the second CDL step following the *HIT* so that the duration ($T_R$) of the input TA pulse is now comprised between $\tau_Q$ and $2\tau_Q$.

Even if the pseudo-period of the pulse train may vary, the TA gain is at first order insensitive to PVT variations if the duration of the pulses remains constant during the circulation process. For this purpose a special care has been taken to ensure equal propagation and transition times for the two edges. Practically, this is easier when the elementary delays are small.

Fig. 6 shows the simulated transfer functions of the TA for various process corners (ss, tt, and ff), voltage (±10% around nominal value) and temperature (−50°C and 65°C) conditions for a theoretical gain value of 8.

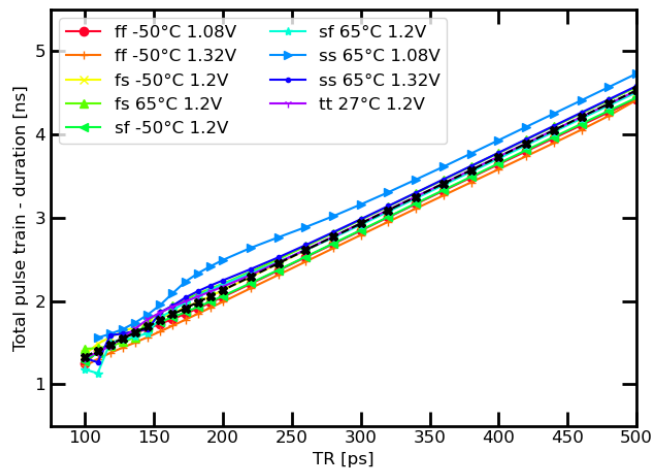

Fig. 6. Simulation of the TA transfer function, for gain $A = 8$ and various processes (ss, tt, and ff), supply voltages, and temperatures (−50 to 65°C).

They have a reasonable linear behavior, with an effective gain close to 8, in the useful range of $T_R$ between 200 ps and 400 ps. To quantify the timing error due to TA non-linearity or gain error, we have subtracted the transfer function of an ideal TA to these transfer functions and normalized the result by the expected gain of 8. These input-referred TA errors are plotted on Fig. 7. In the useful range, for a given case of simulation, this error appears to be nearly constant. This constant term can be seen as an offset that will be canceled during the calibration procedure. In the useful range, the deviation from the constant term is less than ±5 ps for all the cases except the extreme slow case (ss process, 65°C, 1.08V supply voltage) for which it is less than ±10 ps. Thus, for a TDC with LSB = 24.4 ps this will result to an acceptable contribution to its Integral Non Linearity (INL) of less than ±0.5 LSB more likely ± 0.25 LSB).

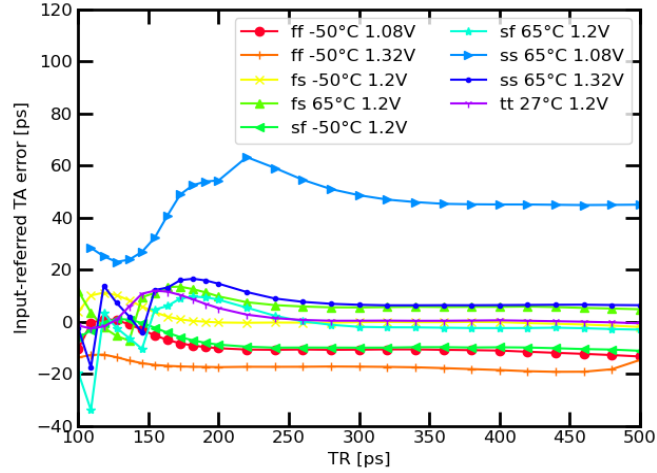

Fig. 7. Input referred-error due to the non-idealities of TA vs input pulse duration $T_R$ for gain $A = 8$ under process (ss, tt, and ff), voltage, and temperature (−50 C–65 C) variations.

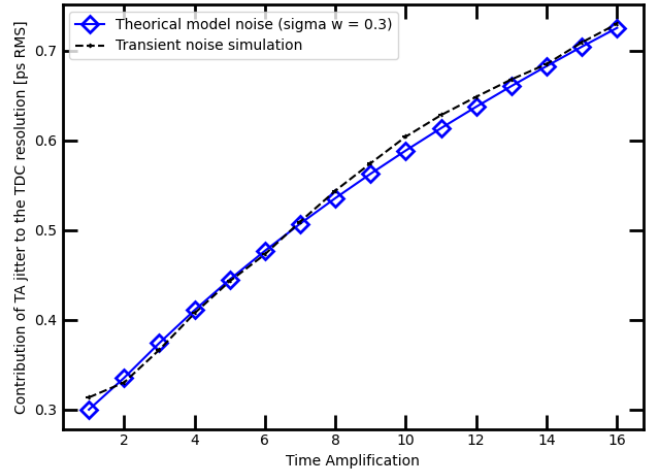

Fig. 8. Input-referred contribution of the TA jitter to the TDC timing resolution. Simulation (blue lozenge) and using (13) (black dots).

The impact of the TA on the timing resolution is studied in the Appendix I. This theoretical approach has been confronted with noise simulations performed with the Spectre simulator [19] to produce the plot shown on Fig. 8. With $\sigma_w = 0.3 \, ps \, RMS$, The simulation results are in very good agreement with equation (13) taken into account a pulse width modulation spread of $\sigma_w = 0.3 \, ps \, RMS$ for each circulation step. This plot shows that even for the time gain of 16, the jitter of only 0.75 ps RMS is added by the TA process, when referred to input of the TA. This is totally negligible compared to the LSB of the TDC, which is 12.2 ps in these conditions.

*3) Fine Delay Line TDC (FTDC)*

The pulse-train duration at the TA output is measured by a FDL. It uses a gated DL (GDL) structure, made of a chain of N identical gated DE (GDE) with a propagation time $\tau_Q$ tunable by a $V_{ctrl\_FTDC}$ voltage. A pulse can propagate in the GDL only when its enable input (*ENGDL*) is activated, whereas the GDL state is frozen when *ENGDL* is off. As shown on the right part of the Fig. 9, this is achieved by cutting or enabling the current path in the starved and standard inverters constituting the GDEs thanks to NMOS and PMOS series-switches (M2, M3, M6 and M7) controlled by the *ENGDL input* and its inverted version *ENGDLb*.

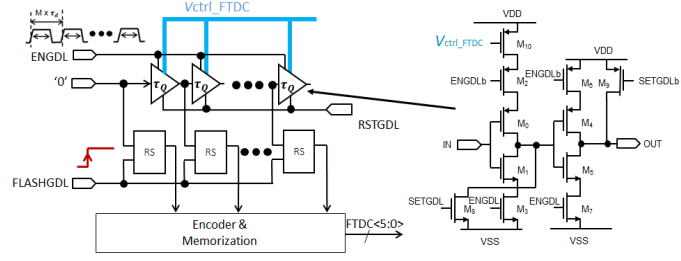

Fig. 9. (Left) Architecture of the FTDC, (Right) schematic of the Gated Delay Element used in the FTDC.

Between consecutive FTDC conversions, the *SETGDL* signal forces all the outputs of the GDEs to '1'. It is released at the beginning of the FTDC conversion, so that the '0' level, set at the input of the first GDE, is ready to propagate in the GDL. This propagation is enabled by the TA output signal, connected to the *ENGDL* input of the GDL. After the alternations of levels '1' and '0' of the pulse train, the final position of the '0'-to-'1' transition in the GDL depends linearly of the "pulse-train duration". It is memorized and encoded, when a *FLASHGDL* signal arrives. This signal, is generated by the TA block after a small delay at the end of the multiplication process.

As with the CTDC, a common servo-controlled master GDL provides the control voltages to stabilize the slave GDLs of the FTDCs embedded in each channel. For simplicity, the delay of the GDL element has the same value as the one of the DL of the CTDC ($\tau_Q$=195 ps). As explained in the previous section, the GDL must cover a time range extended to $A(2.\tau_Q)$. For this purpose, the GDL uses 40 GDEs corresponding to the required length for a TA gain of 16 increased by 8 elements allowing to compensate for time offsets. The GDL coded output is therefore encoded on 6 bits for 4 ($A$=16) or 3 ($A$=8) bits used. Unlike the CTDC, there is no channel-by-channel adjustment of the total duration of the FTDC GDL. Indeed, the impact of an error on the duration of these delay lines will be reduced by the gain of the TA, and are therefore expected to be negligible.

At the end of the conversion, the HIT DFF shown on the Fig. 2 is reset by the *RST-EVENT* signal, preparing the TDC for a new event.

As the asynchronous processing time of the TDC is smaller than 25 ns, the TDC could potentially treat more than one event per FIFO readout clock period. In the prototype chip we have chosen to keep only the first event occurring within a 25 ns clock period, but different options are possible for future applications.

### D. Channel implementation

Fig. 10 shows the block diagram and the physical implementation of a complete TDC channel: the Counter Register captures the MSBs on 8 bits; the CTDC, based on a DL provides 6 intermediate bits (5 of which are significant); and the FTDC, using the time residue amplification method, encodes the LSBs on 5 bits (4 or 3 of which are significant, depending on the configuration). The data from each block are combined to form the final TDC code $D_{out}$ using (8):

$$D_{out} = D_{counter} \times NA - (D_{CTDC} - 1) \times A - D_{FTDC},\qquad(8)$$

where $D_{counter}$, $D_{CTDC}$ and $D_{FTDC}$ are the converted values of the counter, the CTDC and the FTDC respectively. Optionally these data are truncated to 10 bits at the output of this block.

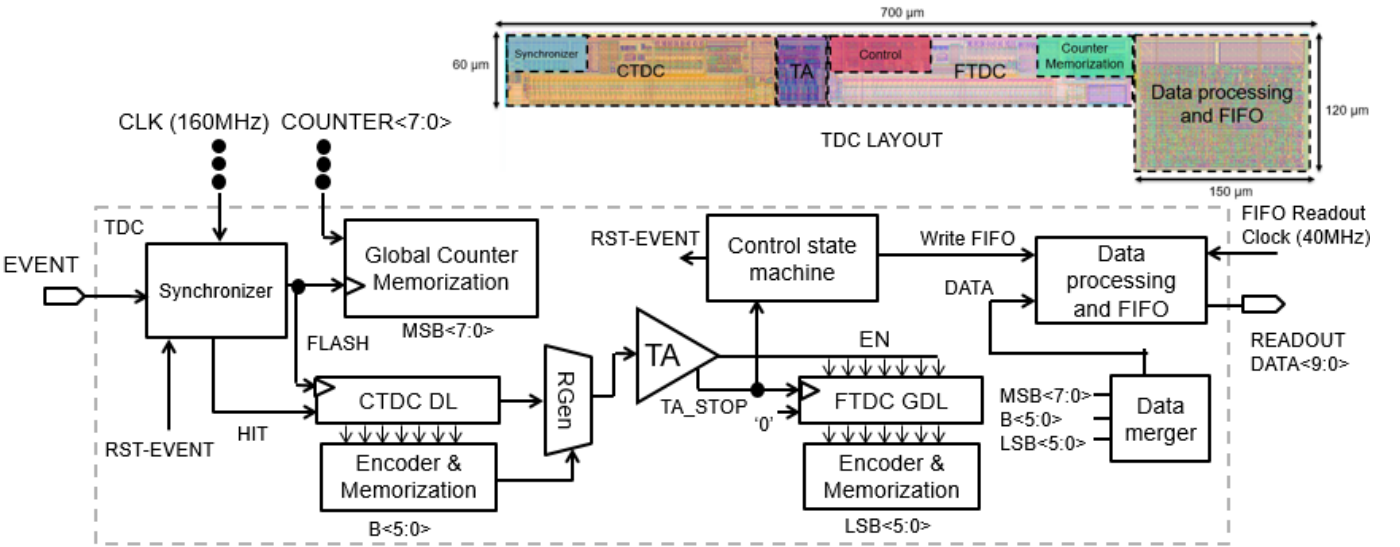

Fig. 10. Block-diagram (bottom) and layout (top) of the proposed TDC channel.

Finally, a control block incorporates state machines that sequence the conversion, store its output data in a FIFO and reset the TDC channel for the next conversion.

The TDC conversion time can be broken down as follows:

- The time between the HIT signal and the STOP of the CTDC, comprised between $4\tau_Q$ and $36\tau_Q$ corresponding to the range from 0.8 ns to 7 ns.
- The time required for the residue extraction and multiplication, defined by $A.T_{loop}$, with $T_{loop}$= 750 ps. Assuming a nominal TA gain of 8, this operation lasts 6 ns. This time also sets the time of use of the GDL. During the same period the CTDC value is encoded.
- Less than 3 ns are then required to encode the FTDC value, to build the final TDC code and to write it in the FIFO.

So that, the conversion is performed within a delay comprised between 9.8 ns and 16.8 ns, smaller than the 25 ns maximum value set by the FIFO readout clock used in the synchronous operation required by the HL-LHC environment. Even for $A$=16, the conversion time is smaller than 23 ns.

## IV. DESCRIPTION OF THE PROTOTYPE AND MEASUREMENT RESULTS

To evaluate the performance of the proposed 3-step TDC structure, a 72-channel TDC prototype was designed and fabricated using a 1.2 V 130 nm digital CMOS process. This section describes the implementation details of this prototype and the results of its characterization measurements.

The layout of a TDC-channel, occupying a 0.051 mm$^2$ surface, is optimized for integration within an electronic front-end channel with a rectangle small-width form factor, as shown on Fig. 10. In operation at its nominal 1.2 V supply voltage, the measured power consumption of a channel is 2.2 mW for the maximum activity corresponding to a hit every 25 ns. This power consumption is reduced to 101 µW if there is no event. The common blocks, including the master DLLs, the 8-bit Gray counter as well as the bias references, the state machines and the FIFO, consumes 1.02 mW. A PLL, locked on a 40 MHz external clock, provides the 160 MHz and 40 MHz cleaned clocks used as time references for the chip. This PLL, based on a ring oscillator structure associated with a counter, will be described in a future paper. Its power consumption is 2.2 mW, and its jitter is better than 9 ps RMS when measured alone with a low jitter reference clock.

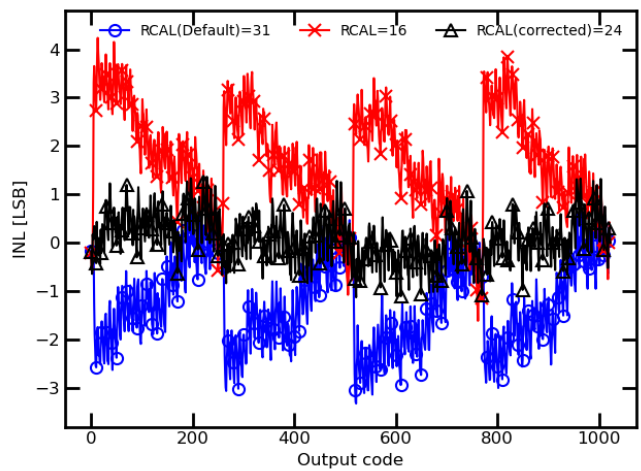

Fig. 11. INLs of a single TDC channel; by $R_{CAL} = 31$ (default) (blue circle), $R_{CAL} = 16$ (red cross) and $R_{CAL} = 24$ (black triangle).

### A. *Measured performance of the TDCs*

Nonlinearities and time resolution are the most important parameters of the TDCs. To measure the nonlinearities of the TDC channels, we have used input signals, asynchronous with the chip clock, generated by the AGC block (shown in the Fig. 1). The obtained results are exactly identical when using external random signals. One million events are recorded to characterize a channel. The Differential Non Linearity (DNL) is obtained using the normalized statistical density of output digital codes provided by the TDC and the INL is calculated by integrating the DNL. All the measurements presented in this paper have been performed in the nominal TDC mode, with the TA gain set to 8, resulting to a LSB of 24.4 ps.

At startup, the control voltage $V_{ctrl}$ provided by the MASTER DLLs (of the coarse and the fine TDCs) are directly fed to slave DLs located in the channels without any correction. The INL measured in these conditions on a typical TDC channel (channel 70) is plotted over the 25 ns range (circle symbols) on Fig. 11. The INL, with a peak-to-peak amplitude of 4 LSBs, is dominated by a pattern with nearly four identical saw teeth, each corresponding to the range of the CDL (6.25 ns). This shape reveals that, as expected, the total CDL propagation time deviates slightly from its expected value. In this particular case, the CDL is actually too fast, for other channels it could be too slow. This can be compensated channel-by-channel using the trimming circuit introduced in III.B.3). Its effect on the slope of the saw teeth is visible on Fig. 11 for two values of the tuning parameter ($R_{CAL}$).

For the optimal tuning (triangle), referred hereafter as HW calibration, the INL is comprised between −1.1 LSB and 1.4 LSB, with a standard deviation of 0.47 LSB RMS (11.5 ps RMS). Even for this optimal tuning, we distinguish 4 zones, corresponding to four 160 MHz periods with 4 slightly different offsets. It is possible to calibrate these offsets and to subtract them from the measurements. This is called hereafter "4-LUT SW calibration" whereas "SW total LUT calibration" stands for a software correction using the full INL pattern.

The corresponding DNL, plotted on Fig. 12, is comprised between −0.7 LSB and 0.9 LSB. As shown on the zoomed view of Fig. 13, this DNL is dominated by a modulo-8 pattern. The source of the pattern has been localized in the FDL implementation that will be improved in the next versions of the chip. Even without this improvement, the measured performance, the absence of missing codes and the seamless connections between the three ranges validate at first order the TDC architecture.

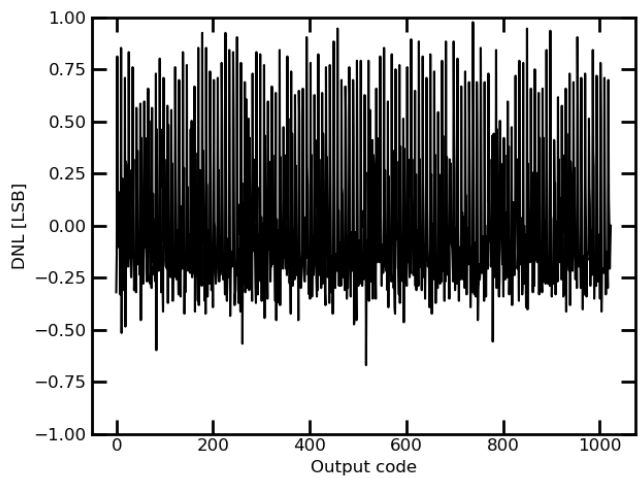

Fig. 12. DNL measured on channel 70, after internal CDL total delay trimming.

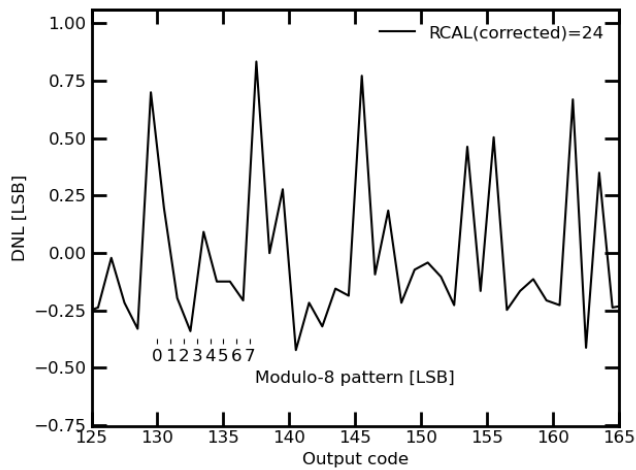

Fig. 13. Zoom on the DNL plot of Fig. 12.

## B. Time interval measurements

To complete the characterization, we have measured the resolution of the time difference between two TDC channels (after HW calibration). For this, two asynchronous signals from an external generator are used, one part is directly sent to a channel whereas the other part is delayed by a cable to feed the second channel. Indeed, the asynchronous test allows to explore the whole phase space between the signal and the clock, including cases likely to cause metastability problems. Practically in our setup, 8 channels from the first TDC subgroup (channels 1 to 36) are simultaneously pulsed with the first signal, whereas the delayed one is pulsing 8 channels from the second TDC subgroup (channels 37 to 72). A typical distribution of the measured time difference between two channels is plotted on Fig. 14. It has a Gaussian shape with a standard deviation of 0.93 LSB corresponding to 22.7 ps RMS. Dividing this value by $\sqrt{2}$, we can deduce a metric of 16 ps RMS timing precision for a single channel. We did not notice any sign of abnormal events in the timing distribution revealing a problem of metastability or range connection.

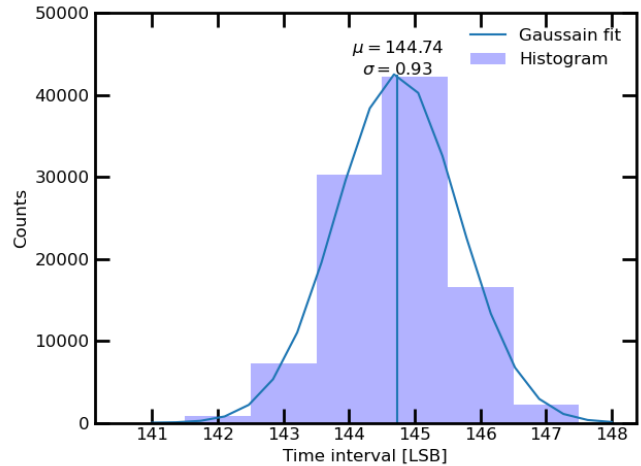

Fig. 14. Distribution of the time difference after CDL trimming (here channel 0 and 70, for a 3.5ns delay)

Using the simultaneous time difference measurements, we can write a set of equations expressing the time difference resolutions as functions of timing precision of individual channels. This $\Delta t$ resolution calculation is presented in Appendix II. The computed values (again with a 3.5 ns inter-pulse delay), plotted on Fig. 15, have a mean value of 14.3 ps RMS and a spread of 1.1 ps RMS, which is consistent with the very first estimation made previously.

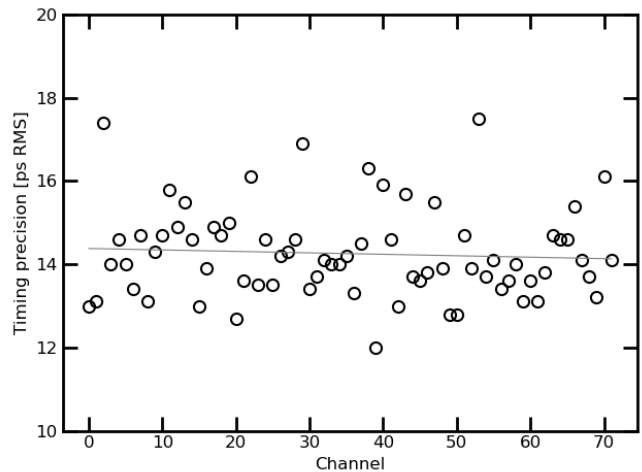

Fig. 15. Timing precision measured for all 72 channels after internal CDL calibration (3.5 ns inter-pulse measurement condition).

To go further, we have measured the time difference precision as a function of the delay value between the input pulses. For this purpose, we used two signals provided by a GFT1004 digital delay signal generator [20], with programmable variable delays, in a range of 0 to 40 ns. The delays between these two signals were precisely measured by an oscilloscope after averaging. For each delay, the channels were pulsed as previously described and the average time difference and timing precision were extracted. The deviations between the measured and expected values are plotted as function of the delay on Fig. 16 for a representative set of channels.

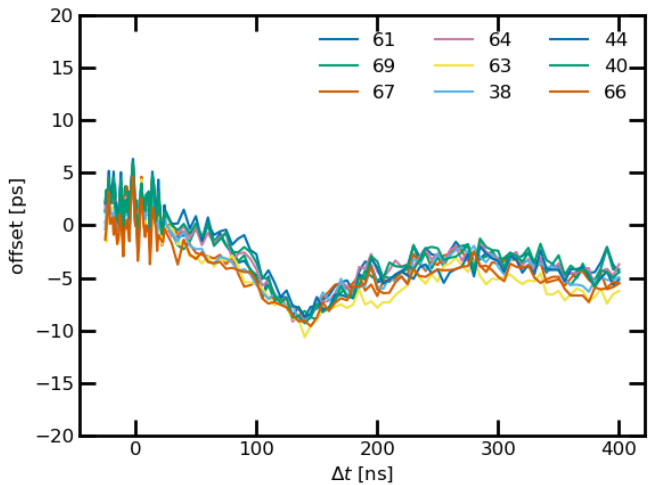

Fig. 16. Deviations between measured and expected time difference for typical channels for increasing time intervals.

For the whole delay range explored, the measured time difference values follow the same trend with less than one LSB deviation from expected values. This pattern is common to all channels and is reproducible. Using the same setup, we have extracted the time precision for each delay in various configurations and plotted it on Fig. 17. Without any calibration, we notice a large pattern with a periodicity corresponding to the 25 ns clock period with a minimum value of 24 ps RMS for zero delay and a maximum value of 40 ps RMS for a long delay.

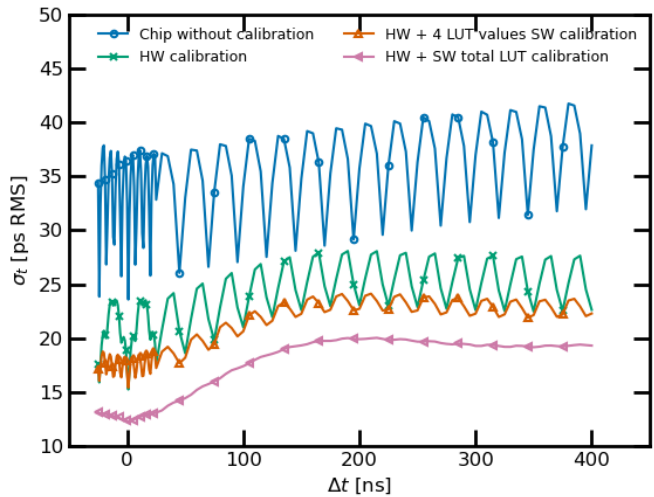

Fig. 17. Time precision for delays between -25 ns and 400 ns, without and with various INL calibrations. The data are average values over 72 channels of a chip.

This is strongly improved after the nominal HW calibration to less than one LSB. The remaining periodic pattern is further reduced after using the "4 LUT SW correction" requiring only 4 constants. This simple correction, requiring very limited resources, permits to obtain a timing precision better than 22 ps RMS, amply satisfying the goal fixed for this design. Applying the "SW total LUT correction" procedure, the periodic pattern of INL is fully canceled and we obtain a timing precision characteristic with a minimal value of 12 ps RMS for small delays increasing up to 18 ps RMS for delays above 200 ns (Fig. 17). In practice, we do not intend to use this complete correction for setups with a large number of channels as it may require substantial resources for its implementation. The "4 LUT SW correction" is a good compromise because it is done in software and can probably be tolerated in many cases. Currently only a few 100kevent/channel taken in few minutes can do this calibration for the entire chip.

Nevertheless we have tried to understand the shape of the fully corrected timing precision.

A dedicated pin of the test chip gives access to the clock generated by the on-chip PLL. We have thus measured the N-cycle jitter of this clock, defined as the standard deviation of the duration of N consecutive clock periods for increasing numbers of periods (and thus increasing time difference $\Delta t$ between the clock edges considered up to 60 μs). Considering that each of the two edges contribute equally to this N-cycle jitter, we can deduce the single edge contribution ($\sigma_t$) by dividing the measured N-cycle jitter value by $\sqrt{2}$. It is plotted on Fig. 18 for two configurations. One uses a nominal FPGA-generated clock as a reference for the PLL. The second configuration uses a low (1 ps RMS) jitter clock as the PLL reference (note that this configuration was foreseen for debug only and as such does not allow to acquire the TDC data). We can first notice that the behavior of the N-cycle jitter is very similar to the one of the time precision plotted also for reference in the same figure, showing an increase for Δt up to 200 ns reaching a plateau. Thus the N-cycle jitter a good candidate to explain the moderate degradation of the timing precision for large delays. We observed a similar behavior with the low jitter reference clock as well, even though the obtained N-cycle jitter is lower. This indicates that, the PLL does not clean the input clock efficiently in a low frequency range and probably adds itself some low frequency jitter.

The values of $\sigma_t$ (5ps RMS) obtained for small N are actually dominated by noise added by our imperfect test setup of the integrated PLL.

Considering these results, we expect an improvement of the timing precision for the large-delay measurements when operating in the final setup with a clean clock. We also conclude that enhancement of the PLL is a key point in order to improve significantly the timing precision of a future chip based on the same architecture.

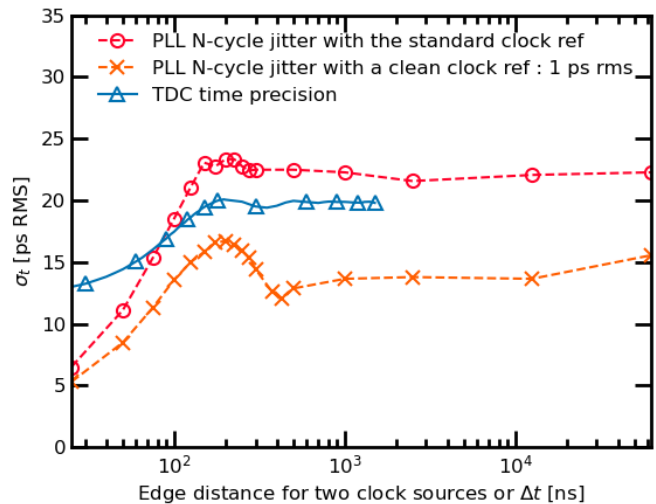

Fig. 18. N-cycle jitter of the embedded PLL as a function of edge distance measured with two clock sources. $\sigma_t$ is the N-cycle jitter divide by sqrt(2) (corresponding to the contribution of one edge). The TDC time precision from Fig. 17 is overlaid for comparison.

## C. *Temperature stability*

Even if for the target applications, the TDCs will operate in an environment with a well-controlled temperature, stability with temperature remains a key point for such a circuit. For instance, we can imagine that local temperature on the silicon could vary, depending on the chip or board activity. In this design, the variations of the key timing parameters are supposed to be controlled by the master DLLs with respect to the temperature changes. But as the design relies on the stability of the matching between these master elements and open-loop blocks located in the channels, it is important to check that the TDC channel performance is not deteriorated when the temperature deviates from the value used for calibration. This was checked in the −35°C to 65°C range using a climatic chamber.

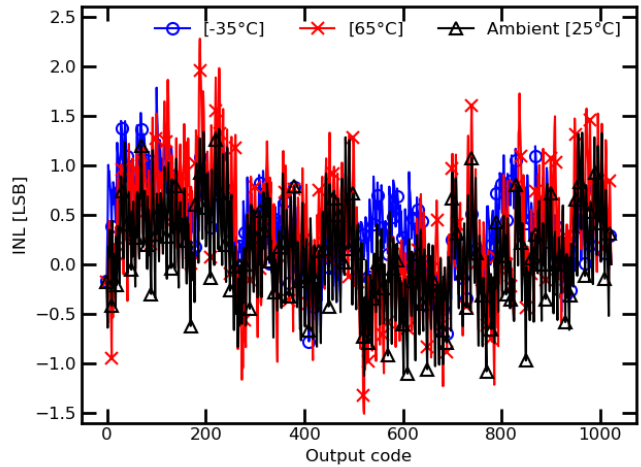

Fig. 19. INL of a TDC channel#70 after CDL delay calibration ($R_{CAL} = 24$) at room temperature 25°C (black triangles), at 65°C (red crosses) and -35°C (blue circles).

After a HW calibration achieved at 25°C, we performed measurements within the temperature range from −35°C to 65°C. The measured INL, shown on Fig. 19, only changes slightly and remains in the −1.5/2.2 LSB range. The standard deviation of the INL rises with temperature from 10.5 ps RMS at −35°C to 15 ps RMS at 65°C. As previously mentioned, the value at room temperature is 11.5 ps RMS.

In the same conditions, we also measured the timing precision, as described in the previous section, for pulses delayed in steps of 3.5 ns and for the three temperatures. The results plotted on Fig. 20 reveal a noticeable performance improvement at low temperature with time precision of 10.1 ps RMS at −35°C. Non-surprisingly, the timing precision follows the evolution of the INL with temperature.

These measurements show a good stability of the TDC performance despite of the wide temperature range. These measurements associated with the non-linearity and timing precision measurements meet our initial requirements and validate our architectural choices.

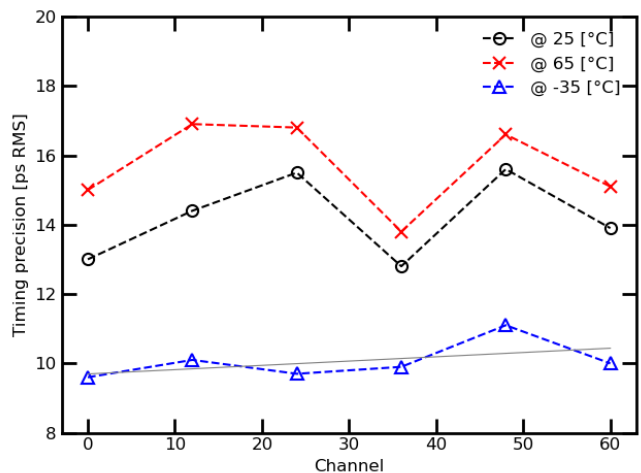

Fig. 20. Timing precision at different temperatures of six TDC channels after internal CDL trimming; at room temperature 25°C (black cross), 65°C (red circle) and -35°C (blue triangle) for pulses delayed by 3.5 ns .

## V. CONCLUSION

A novel low-power multichannel TDC based on a 3-step structure using DLLs associated to the Time Residue Amplification technique has been implemented and successfully tested. In its nominal mode, it is able to perform a time conversion in less than 25 ns for a 10-bit range with a LSB of 24.4 ps. It operates with a 1.2 V supply and consumes between 101 μW and 2.2 mW per channel depending on the hit rate. A special care has been taken in the design to ensure stability with PVT variation. This includes the use of servo-control by Master-DLL and possibility to trim each channel to compensate for mismatches. Once trimmed using a very simple procedure, all TDC channels exhibit excellent INL (between −1.1 LSB and 1.4 LSB) and DNL (better than 1 LSB).

The time precision for asynchronous signals is better than 25 ps RMS without any correction, and can be reduced to 22 ps after a very simple offline correction. As described, the performance can be improved further by more advanced calibrations and with a better clock reference.

Moreover, performance is stable when the temperature is changed. A time precision as good as 10 ps RMS was measured for a −35°C temperature. This work validates our new TDC architecture that is currently being integrated in three different ASICs, with minor improvements. It should be noted that the on-chip programmable asynchronous clock generator ACG allows an autonomous *in-situ* calibration of the chip as well as an individual adjustment of the channel offsets during the chip operation.

### APPENDIX I: PULSE-TRAIN WIDTH MODULATION DUE TO RECIRCULATION IN THE PULSE REPLICATOR

While the pulses recirculate in the TA during the multiplication process, the jitter of the elementary digital delays contributes to degradation of the information about the width of the original pulse [17]-[18]. The problem is not as simple as the jitter accumulation during edge propagation but is related to the dispersion of the width of the pulses of the train. To evaluate the effect, we have to take into account the jitter of both the rising and falling edges. For this, we make the hypothesis that each circulation $i$ modulates the pulse width with a standard deviation $\sigma_{wi}$, and that these modulations are uncorrelated between two consecutive revolutions (hypothesis of high frequency jitter). The standard deviation of the width of the 2nd pulse (generated after two circulations) is given by:

$$\sigma_{w2nd\ pulse}{}^2 = \sigma_{w1}{}^2 + \sigma_{w2}{}^2, \qquad (9)$$

but the standard deviation of the pulse-train duration after two circulations is :

$$\sigma_{train,2}{}^2 = (2.\sigma_{w1})^2 + \sigma_{w2}{}^2, \qquad (10)$$

as the two terms related to the first circulation are fully correlated. Using, the same analysis, iteratively, we can calculate the variance of the total width of the output pulse train after n circulations:

$$\sigma_{train,n}{}^2 = \sum_{i=1}^{n} (i.\sigma_{wi})^2 = \sum_{i=1}^{n} (i.\sigma_w)^2 \qquad (11)$$

considering that the magnitude of the pulse width modulation has a constant value, $\sigma_w$, for all the circulation steps. This gives a standard deviation of the total width of the amplified pulse of:

$$\sigma_{train,n} = \sigma_w . \sqrt{\frac{n.(n+1).(2n+1)}{6}} . \qquad (12)$$

By dividing it by the TA gain, we obtain the input-referred standard deviation on the TDC time measurement due to the TA jitter:

$$\sigma_{in,n} = \sigma_w . \sqrt{\frac{(n+1).(2n+1)}{6.n}}, \qquad (13)$$

### APPENDIX II: $\Delta t$ RESOLUTION

The timing resolution $\sigma_i$ for each channel $i$ is extracted by measuring the standard deviation $\sigma_{ij}$ of distribution of the time difference between channels $i$ and $j$. Assuming that the time measurements are not correlated in-between channels, we have for $i \neq j$:

$$\sigma_{ij}^2 = \sigma_i^2 + \sigma_j^2 . \qquad (14)$$

For $N$ channels, this gives a set of $(N-1) \times N/2$ independent equations for $N$ unknown $\sigma_i$ which can be solved for $N \geq 3$ (and is over-constrained for $N > 3$). For convenience, we introduce the variable $\sigma_t^2 \equiv \sum_i \sigma_i^2$. We then consider a subset of $N+1$ equations with $N+1$ unknowns $\sigma_i^2$ for $i \in [1,N]$ and $\sigma_t^2$) formed by summing over $j \neq i$ equation (14). Thus, this subset of equations is given by:

$$
\begin{cases}
\sigma_t^2 = \displaystyle\sum_j \sigma_j^2 \,, \\[2mm]
(N-2)\,\sigma_i^2 + \sigma_t^2 = \displaystyle\sum_{j\neq i} \sigma_{ij}^2 \quad \text{for } i \in [1,N]\,,
\end{cases}
\tag{15}
$$

This system admits as solution:

$$
\sigma_t^2 = \frac{1}{2N-2} \sum_{i,j\neq i} \sigma_{ij}^2 \,,
$$

$$
\sigma_i^2 = \frac{1}{N-2}\left[\left(\sum_{j\neq i} \sigma_{ij}^2\right) - \sigma_t^2\right]\,.
\tag{16}
$$

## ACKNOWLEDGMENT

The authors would like to thank the CMS HGCAL - Phase II Calorimeter Endcaps (CE) community for their helpful discussions and suggestions regarding this work. The authors also would like to thank the LLR (Laboratoire Leprince-Ringuet) and the CERN engineers for software and firmware implementation.